\documentclass[aps,prd,superscriptaddress,nofootinbib,showpacs,showkeys]{revtex4}
\usepackage[cp1251]{inputenc}

\usepackage{amsmath,amsfonts,amssymb,amsthm}           % Basic AMSTeX packages
\usepackage{dsfont,textgreek}                          % Double-stroke font and text-mode Greek letters
\usepackage[bb=boondox]{mathalfa}                      % Cool double-stroke fonts, including zero
\usepackage{graphicx}                        % Graphics and floating figure support
\usepackage{epsfig}
\usepackage{multirow}
\usepackage{epstopdf}                                 % Allow immediate EPS -> PDF conversion when PDFTeXified (instead of TeXify + DVI2PDF)
\usepackage{verbatim}

\newcommand{\idMatrix}{\mathds{1}}
\newcommand{\pd}{\partial}
\newcommand{\ii}{\mathrm{i}}
\newcommand{\const}{\mathrm{const}}
\newcommand{\diff}{\mathrm{d}}

\newcommand{\Angstrom}{\text{\AA}}
\newcommand{\Reals}{\mathbb{R}}
\DeclareMathOperator{\diag}{\mathrm{diag}}   % diagonal matrix
\newcommand{\hc}{^\dagger}               % Hermitian conjugate
\newcommand{\cc}{^\ast}               % Hermitian conjugate
\newcommand{\bvec}[1]{\mathbf{#1}}                     % 3-dimensional vector typed in bold
\newcommand{\twoCol}[2]{\begin{pmatrix} #1 \\ #2 \end{pmatrix}}

\newcommand{\twoMatrix}[4]{ \begin{pmatrix}{#1} & {#2} \\ {#3} & {#4} \end{pmatrix} }

\newcommand{\IV}{$I$--$V$}                            % I-V curve
\newcommand{\VI}{$V$--$I$}                            % V-I curve

\begin{document}

\title{Stability of quantized conductance levels in memristors with copper filaments:
toward understanding the mechanisms of resistive switching}
% beta version: Stability of quantized conductance levels in memristors with copper filaments: an experimental-theoretical study
% A possible version: "with copper electrodes and its possible driving factors"
% older version: ... in parylene-based memristors...
% old: Stability of quantized conductance levels in parylene-based memristors with copper electrodes: an experimental-theoretical study

\author{Oleg~G.~Kharlanov} \email{kharlanov@physics.msu.ru}
\affiliation{Faculty of Physics, Lomonosov Moscow State University, % 1/2 Leninskie Gory,
119991 Moscow, Russia}

\author{Boris~S.~Shvetsov}
\affiliation{National Research Center ``Kurchatov Institute'', 123182 Moscow, Russia}

\author{Vladimir~V.~Rylkov}
\affiliation{National Research Center ``Kurchatov Institute'', 123182 Moscow, Russia}%
%\affiliation{Kotelnikov Institute of Radio Engineering and Electronics RAS, 141190 Fryazino, Moscow Region, Russia}

\author{Anton~A.~Minnekhanov} \email{minnekhanov\_aa@nrcki.ru}
\affiliation{National Research Center ``Kurchatov Institute'', 123182 Moscow, Russia}

%-----------------------------------------------------------------------------------
% PACS codes:
% 03.65.Nk QM::scattering theory
% 05.30.Fk St-Nlin::Fermion systems and electron gas
% 05.60.Gg St-Nlin::Quantum transport
% 34.70.+e AtomicCollisions::Charge transfer
% 62.23.Hj MechCondMat::Nanowires
% 62.23.Pq MechCondMat::Composites (nanosystems embedded in a larger structure)
% 66.30.Qa MechCondMat::Electromigration
% 72.10.-d ElectronicTransport::Theory of electronic transport; scattering mechanisms
% 73.63.-b ElTrans_LowDim::Electronic transport in nanoscale materials and structures
% 81.07.Vb MatSci::Quantum wires
% 84.37.+q Electronics::Measurements in electric variables (including V, I, R, C, L, Z, etc.)
% 85.35.-p Electronics_Devices::Nanoelectronic devices
%-----------------------------------------------------------------------------------


\begin{abstract}
    Memristors are among the most promising elements for modern microelectronics, having unique properties such as quasi-continuous change of conductance and long-term storage of resistive states. However, identifying the physical mechanisms of resistive switching and evolution of conductive filaments in such structures still remains a major challenge. In this work, aiming at a better understanding of these phenomena, we experimentally investigate an unusual effect of enhanced conductive filament stability in memristors with copper filaments under the applied voltage and present a simplified theoretical model of the effect of a quantum current through a filament on its shape. Our semi-quantitative, continuous model predicts, indeed, that for a thin filament, the ``quantum pressure'' exerted on its walls by the recoil of charge carriers can well compete with the surface tension and crucially affect the evolution of the filament profile at the voltages around 1~V. At lower voltages, the quantum pressure is expected to provide extra stability to the filaments supporting quantized conductance, which we also reveal experimentally using a novel methodology focusing on retention statistics. Our results indicate that the recoil effects could potentially be important for resistive switching in memristive devices with metallic filaments and that taking them into account in rational design of memristors could help achieve their better retention and plasticity characteristics.
\end{abstract}

    \maketitle

    \section{Introduction}
    \label{sec:Intro}
    Memristors as new representatives of the microelectronics element base have been actively studied since 2008, {when a manufactured $\text{TiO}_2$-based device was claimed to be a physical implementation of the missing fourth circuit element type---along with the resistor, the capacitor, and the inductor~\mbox{\cite{ref1}}. In fact, the picture describing the four idealized circuit elements was introduced decades earlier~\mbox{\cite{ref1_Chua}}, while the actual memristors used today have extended the original concept and absorbed a long history of studies of various similar devices and the associated physical effects~\mbox{\cite{ref7, ref9.3, refKozicki, refAono, refKund, refKhvalkovskiy}}.} With such advantages as the ability to set and maintain the
    necessary resistive state, quasi-continuity of the conductance, scalability, and low power consumption needed for resistive
    switching (RS), memristors are now viewed as a promising route towards creation of resistive random-access memory~\cite{ref2}, in-memory computing
    systems~\cite{ref3}, and hardware neural networks~\cite{ref4, ref5, ref6}. In the latter case, they can play a role of both
    synapses and components of neurons~\cite{ref7,ref8}, eliminating the need to use dozens of transistors, which greatly simplifies the circuitry~\cite{ref9}. Importantly, a large number of accessible stable resistive states (a property also referred to as the plasticity) greatly enhances the versatility of memristors in such neuromorphic devices~\cite{ref9.1}. Thus, understanding of the physical mechanisms
    underlying RS and retention of resistive states in various types of memristors is required to fuel the progress of the
    technology towards the new applications~\cite{ref9, ref5, ref9.2}. In fact, the
    corresponding physics remains quite intricate, depending on the particular memristor type, thus, combining experimental and theoretical efforts appears a fruitful strategy for unraveling it.

    Many materials allowing for RS, along with the possible RS mechanisms, have been proposed~\cite{ref7}, and among the latter, two mechanisms involving ion migration, which are quite similar to those important for the functioning of biological synapses, can be noted here. These are the mechanism of electrochemical metallization (ECM)~\cite{ref4,ref10,ref11}, with metallic bridges formed in a dielectric, and the valence change mechanism (VCM)~\cite{ref12}, assuming formation of conductive filaments from oxygen vacancies. Memristors based on these RS mechanisms typically have a sandwich structure with the metal electrodes either active, such as copper or silver, or inert, such as platinum, between which lies a dielectric layer~\cite{ref10}. The filaments---metallic in ECM and oxygen-vacancy ones in VCM memristors---cause RS by growing and short-circuiting the top and bottom electrodes, when a voltage pulse is applied to the structure, and switching the device to its low-resistance state (LRS, or the on-state)~\cite{ref10,ref11}.

    Conversely, when a negative voltage pulse is applied, these
    filaments may shrink in size or even rupture, leading to a decrease in conductance and switching the memristor back to its high-resistance
    state (HRS, or the off-state). During the switching process, individual filaments of nanometer diameter can form,
    in which quasi-1D ballistic transport takes place, i.e., electrons pass through them without scattering~\cite{ref13}.
    In this case, one can observe the so-called conductance quantization effect, in which the conductance of the device changes
    stepwise with the growing filament, following the Landauer formula~\cite{ref14}:
    \begin{equation}\label{G_quantization}
        G = \frac{2 n e^2}{h} = n G_0, \quad n = 0, 1, 2, 3, \ldots,
    \end{equation}
    where $e$ is the electron charge, $h$ is the Planck's constant, and $G_0 = 2 e^2 / h \approx 77.5~\text{\textmu{}S}$ is the conductance quantum, the factor of two coming from the twofold spin degeneracy. In fact, half-integer conductance plateaus, proportional to $G_0/2$ and formally corresponding to $n = 0.5,\, 1.5,\, 2.5, \ldots$ in Eq.~\eqref{G_quantization}, have also been observed experimentally~\cite{ref13,ref16}. Moreover, memristors, unlike, for example, the 2D electron gas~\cite{ref17}, are remarkable in that conductance quantization (both
    integer and half-integer) can be well observable even at room temperature~\cite{ref13,ref18,ref19}, which could be caused, among other things, by the stabilizing function of the dielectric medium surrounding the conducting filament. Last but not least, the ``precision'' of conductance quantization in ECM and VCM memristors is one of the favorable factors for their high plasticity.

    Note that a single conductance quantum, theoretically, corresponds to one fully open conductive channel for electrons~\cite{ref14, ref15}, i.e., to the transmission probability $\mathcal{T} = 1$. For a general filament geometry, however, the transmission probabilities for the conductive channels in it do not have to equal either zero (closed channels) or unity (open, ballistic ones), especially, when it comes to the narrowest segments of thin filaments that are only several atoms thick~\cite{ref14, Buttiker, Sablikov}. Thus, clearly discernible quantized-conductance states in (certain types of ECM/VCM) memristors indicate the existence of a physical mechanism directing creation or evolution of filaments towards those with (half-)integer $G$ values and/or enhancing the stability of the latter ones. In other words, the ``shape'' of the filament is expected to affect its fate, and the connection between the two, among other things, could have to do with the transmission probabilities, which are quantum in nature.

    A possible relation between the stability of a conductive filament and the conductance (virtually, the filament geometry), as well as its role in the emergence of the conductance quantization effect, also appears interesting in the context of experimental observations in different types of memristors. For example, half-integer quantization is not always observed in ECM memristors, but much more often in VCM ones
    %\alert{, despite the filament thicknesses are typically considerably greater in the latter ones}
    (see Ref.~\cite{ref13} and references therein). At the same time, we have recently shown that room-temperature half-integer quantization is well observed in parylene [poly(para-xylylene), PPX] based memristors with copper electrodes~\cite{ref16,ref19}. Such memristors consist of a thin dielectric layer of PPX between the bottom indium tin oxide (ITO) and the top copper electrode, and are very promising due to their good memristive properties, along with the biocompatibility of PPX~\cite{ref20,ref21}. The Cu filament formed during their resistive switching can be reduced in diameter to the atomic scale, leading to occurrence of conductance levels according to Eq.~\eqref{G_quantization}, despite the considerable stochasticity of the filament dynamics. Half-integer quantization is also observed in other memristors with copper~\cite{quantization_copper1, quantization_copper2} and silver filaments~\cite{quantization_silver1}, etc. Thus, it is highly desirable to uncover the microscopic laws or trends of the formation and lifecycle of filaments with quantized conductance values.

    On the theoretical side of the issue, recent progress in quantum transport simulations driven by density-functional and nonequilibrium Green's function methods enables one to calculate the transmission probabilities for a given (moderate) number of filament atoms with a prescribed geometry
    (see, e.g., Refs.~\cite{Krishnan_QCond_JJAP2017, Khomyakov_PRB2005, Bell_CPC2015, Long_Qwires_APL2013}) or to perform a molecular-dynamics simulation of
    the filament evolution or ion migration~\cite{WangIelmini_NatComm2019, Clima_AIMD_APL2012}. However, the effect of the applied voltage and/or the flowing quantum current on the stochastic evolution of the conductive filament represents a challenge, at least because of the required computational resources and
    a variety of initial geometries. Thus, the very theoretical explanation of how a filament in electric field (statistically often) takes on a shape consistent with conductance quantization~\eqref{G_quantization} is a question to be settled, as well as the question of the enhanced stability of filaments with a quantized conductance.

    In view of the importance of the stability issue and the time evolution of different conductive filaments formed in a memristor, in this work we analyze the retention statistics of Cu/PPX/ITO memristors for different reading voltages and complement this experimental study with a theoretical discussion of a simplified continuous model of the filament shape evolution under the quantum current. Though such a model is able to provide only a semiquantitative insight into the effects taking place, it predicts quite a counterintuitive stabilization effect of the current on the filament \emph{for both current directions}, which is also observed in our experiment. {Interestingly}, our model also predicts considerable changes in the geometries of thin filaments due to electron recoil effects---a kind of quantum counterpart of electromigration~\cite{Electromigration1,Electromigration2}---for quite realistic voltages around 1--2~V, typical of RS in memristors~\cite{ref13}. {Moreover, it turns out that even at ten times smaller reading voltages used in our retention experiments, the additional current-induced stability is predicted to be at the level of roughly $kT$ per one Cu atom (in energetic terms), potentially affecting ion migration.} Thus, we suggest that the experimentally observed effect of current-enhanced filament stability could indeed be rooted in the {quantum regime of charge-carrier recoil}. {In this context, it should be noted that other effects of the electric field or current on the conductance (or conductivity)  are known, e.g., the instabilities due to negative differential resistance, which can lead to formation of conductive domains/filaments~{\cite{Ridley}}, and the electric-field-driven metal-insulator phase transitions~\mbox{\cite{Han,Chiriaco}}. Our study, in contrast, sheds light on the conductive filament geometry on the nanoscale and suggests a physically transparent bridge between the quantized conductance of such filaments (and the associated transmission coefficients) and the stability of the filament profile. Therefore}, by no means attempting to describe the whole microscopic side of the RS effect in this study, we hope that our model describes a mechanism at work in memristive devices and after further elaboration could catch important trends of the conductive filament dynamics. Moreover, our study highlights the potential of the retention statistics analysis as a non-disturbing probe of the physics underlying filament evolution: such a technique keeps the filament intact for a considerable time, working well below the RS voltage. To the best of our knowledge, this is the first application of such a technique to memristive devices.

    Accordingly, the paper is organized as follows. We start from a description of our experimental methodology and results on the conductance quantization in Cu/PPX/ITO memristors in Sec.~\ref{sec:Results_quantization}, revealing a clear dependence of the retention on the resistive state. This experimental indication, also supported by a preliminary, semiqualitative theoretical analysis, inspires a further experimental investigation of the retention statistics (Sec.~\ref{sec:Results_stability}). The theoretical argument, along with a number of formulas it is based on, follows in Sec.~\ref{sec:Results_motivation}; finally, we present an in-depth numerical analysis of the current-enhanced filament stability within our model in Sec.~\ref{sec:Results_simulation}. Note, however, that the research reported did not follow such a linear trajectory from the experiment to the theory or vice versa, but rather was a synergy of the two; the order chosen in Sec.~\ref{sec:Results} is rather arbitrary. We have also moved several computational details and semiquantitative arguments to the Appendix to make Sec.~\ref{sec:Results} more readable and focused. Finally, we discuss the overall picture and the possible outlook in Sec.~\ref{sec:Conclusion}. Details of the experimental techniques used to obtain and study the samples are outlined in Sec.~\ref{sec:Methods}.

    %=================================================================================================================================
    \section{Results and discussion}
    \label{sec:Results}

    \subsection{Performance of Cu/PPX/ITO memristors and conductance quantization in them}
    \label{sec:Results_quantization}

\begin{figure}[ht]
        \includegraphics[width=17cm]{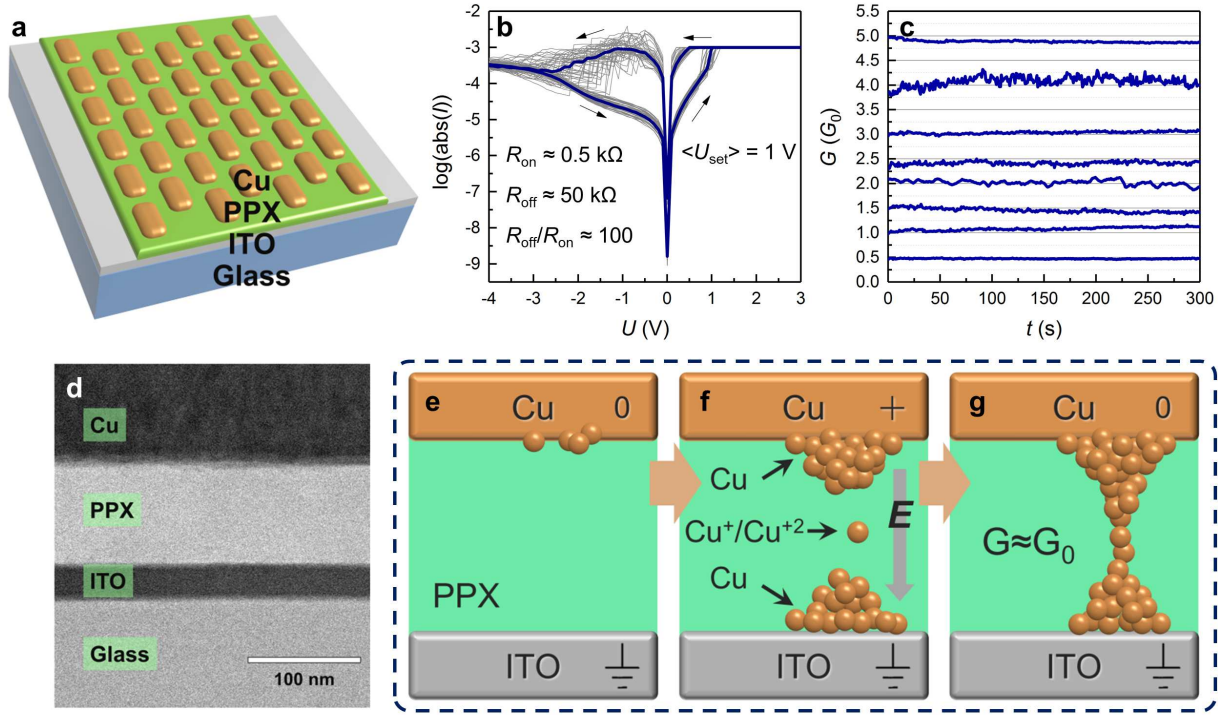}
        \caption{Properties of Cu/PPX/ITO memristors. (a) A sketch of an array of memristive elements. (b) Typical \IV~curves of the memristors, 50 cycles. The median curve is highlighted in dark blue. (c) Demonstration of a multilevel character of RS in memristors and storage of the specified intermediate resistive states.
        %In this figure, the states whose conductance is a multiple of the conductance quantum $G_0$, are presented.
        (d) {Transmission electron microscope (TEM)} image of a cross-section of the memristor with a clearly visible sandwich structure.  (e)--(g) Schematic representation of the cross-sectional area of the memristor, where a conductive filament is formed: {copper atoms at the irregularities of the Cu-PPX boundary, shown on the top electrode in panel~(e), undergo oxidation} under the positive voltage applied to the top electrode (f), which is followed by the movement of the ions to the bottom electrode; here, {Cu ions} are reduced and form a growing filament, which finally connects with its top part and forms a conductive bridge (g).}
        \label{fig:samples}
    \end{figure}

    Let us first discuss general electrophysical characteristics of the Cu/PPX/ITO memristive structures we were using in this study: a schematic representation of such a structure is presented in Fig.~\ref{fig:samples}(a). The samples demonstrated a stable RS effect with a signature hysteresis loop observed in the \IV~curves [Fig.~\ref{fig:samples}(b)]. The 50 curves presented in Fig.~\ref{fig:samples}(b) were recorded consecutively at a sweeping rate of $1~\text{V/s}$ in $100~\text{ms}$ steps; {in all our measurements discussed hereinafter, the voltage was applied to the top electrode (Cu), while the bottom electrode (ITO) was grounded, as indicated schematically in {Fig.~\ref{fig:samples}}(e-g)}. Fig.~\ref{fig:samples}(b) also shows the median \IV~curve constructed by combining the calculated median values from 50 current measurements per each voltage point. The memristors studied had a low cycle-to-cycle voltage variation for the HRS-to-LRS switching (namely, for RS from $R_{\text{off}} \approx 50~\text{k\textOmega}$ to $R_{\text{on}} \approx 500~\text{\textOmega}$): the mean set voltage value for 50~cycles was $U_{\text{set}} = 1.0\pm0.1 ~\text{V}$. It should also be noted that our memristors did not require forming, i.e., in our case, a specific first switching from the pristine state of a manufactured device to the LRS, usually performed at approximately the same voltages as those for the subsequent RSs.

    In general, the observed RS can be characterized as well reproducible. Furthermore, this result is consistent with our previous works on PPX-based memristors~\cite{ref19,ref16}, which indicates a high level of reproducibility of the electrophysical parameters of the structures in the process of synthesis. This is also confirmed by the high plasticity of the studied PPX memristors: their resistance, as before, can take on intermediate values in the $(R_{\text{on}}, R_{\text{off}})$ range and remains stable over time. Some of such resistive states are shown in Fig.~\ref{fig:samples}(c), where we, referring to a more detailed discussion in Sec.~\ref{sec:Results_stability}, demonstrate the time evolution of the conductance measured in units of the conductance quantum $G_0$. The resistive states shown in this figure were set using the ``write-verify'' algorithm with a precision of $0.1~G_0$. {This algorithm, developed by us earlier~{\cite{WriteVerify}}, is based on a repeated application of voltage pulses to the memristor in order to set up a desired resistive state with a high accuracy. Namely, at the $k$th step of the loop, a verification operation is first performed, revealing whether the conductance lies within the target range (say, between $4.9$ and $5.1~G_0$); then, if is does not, a ``write'' operation follows, trying to adjust the conductance up or down, the amplitude of the writing pulses growing with $k$. Further details of the ``write-verify'' algorithm and its applications can be found in Ref.~{\cite{WriteVerify}}.}
    %{Note that the external contribution to the resistance of the entire system (including contacts, wires, etc.) can be neglected, because it is no more than $100~\text{\textOmega}$ and even for the $5~G_0$ level ($2600~\text{\textOmega}$) the inaccuracy is no more than 4\%.}

    We have previously found that RS in PPX-based memristors appears to be driven by electromigration of metal ions (copper in our case) from the top electrode to the bottom one, with the formation of metallic bridges~\cite{ref16}. In other words, these structures are ECM-type memristors. The memristors studied here showed practically no difference from Ref.~\cite{ref16} in their \IV~curves and the microscopic structure [Fig.~\ref{fig:samples}(d)], so we also assign them to the ECM class. Thus, if one forms a conducting bridge relatively accurately during the set process, i.e., extremely slowly and with a negative ``conductance-voltage'' feedback (see below), it is possible to obtain a filament only a few atoms wide in its narrowest part. This process is shown schematically in Fig.~\ref{fig:samples}(e--g). In this narrowest segment of the filament, ballistic conduction can be observed, according to Eq.~\eqref{G_quantization}, with electrons passing through without scattering.

\begin{figure}[ht]
        \includegraphics[width=17cm]{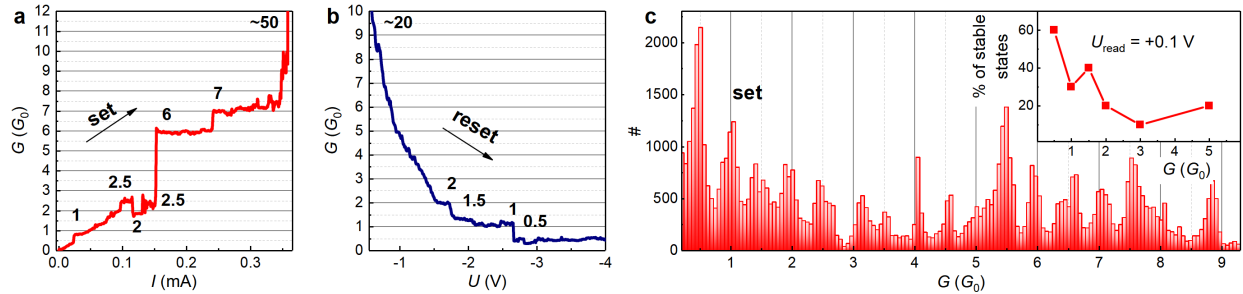}
        \caption{Conductance quantization in Cu/PPX/ITO memristors. (a) An example \VI~curve of an RS of the memristor to the LRS (a ``set'' process), the sweeping rate is $5\times 10^{-7}~\text{A/s}$. (b) An example \IV~curve of a reverse RS of the memristor to the HRS (a ``reset'' process), the sweeping rate is $0.01~\text{V/s}$. (c) A histogram of the conductance values obtained by processing 10~``set'' switches, about $10^5$ data points in total. The inset shows the percentage of stable states amongst all the observed states with a given quantized conductance $G$ ({from a series of 10 measurements for each $G$ point,} see details in the text): higher-$G$ states are statistically less stable for a fixed reading voltage $U_{\text{read}} = +0.1\text{ V}$. In panels (a) and (b), the measured values of the voltage and the current, respectively, are converted into the conductance $G$ for a better visualization.}
        \label{fig:set-reset-quants}
    \end{figure}

    The mentioned negative feedback is an important detail in obtaining observable quantized conductance states. In fact, the very effect of conductance quantization in PPX-based memristors was detected by us unintentionally~\cite{ref16} during a detailed study of the stepwise segment of the \IV~curves for the RS process of the structure back to the HRS. It turned out that the conductance of the structure in this case lies statistically more often near the multiples of $G_0$. At the same time, during direct switchings to the LRS, the quantized conductance levels were not observed even at relatively low voltage sweeping rates. Further it became clear that the reason for the latter was a positive ``conductance-current'' feedback: when under a fixed voltage, an RS occurs at a certain point of the \IV~curve, increasing the conductance, the current through the structure also jumps up immediately after that. This, in turn, pushes the growth of the filament cross-section, which results in a further avalanche-like growth of the current. At the same time, recording the current-voltage characteristics in the current source mode, i.e., measuring the dependences of the voltage on the current (the \VI~curves), we turn this feedback into a negative ``conductance-voltage'' one: if at some fixed value of the current, an RS occurs with the conductance jumping up, this has a negative effect on the voltage~\cite{ref16, ref19}. The latter factor slows down the switching process and the filament growth, which allows us to observe stable conductance levels at the multiples of $G_0$. In contrast, when switching to the HRS, the negative feedback is realized in the voltage source mode (i.e., when measuring the current against the voltage, or the \IV~curve). That is why previously we observed conductance quantization only at reset switches~\cite{ref16}. Of course, in order to register the conductance quantization at cyclic switching, a low sweeping rate is also a necessary condition.  Specifically, we experimentally determined the following optimal values: $\sim 0.01~\text{V/s}$ for voltage sweeps and $\sim 5\times 10^{-7}~\text{A/s}$ for current sweeps. Examples of \VI~and \IV~curves obtained by the methodology described above are shown in Fig.~\ref{fig:set-reset-quants}(a,~b). It can be clearly seen that the conductance takes on not only integer, but also half-integer values in the units of $G_0$. This is even more evident in the histograms of conductance values obtained by repeatedly recording the \IV~curves. For example, Fig.~\ref{fig:set-reset-quants}(c) shows a histogram for 10~switches to the LRS, plotted for about $10^5$ data points in total. {The (less frequently observed) conductance values between integer and half-integer levels are a result of the noise and stochasticity: they can arise, e.g., from thermal fluctuations of the filament geometry (note that integer conductance is predicted by the Landauer formula~{\eqref{G_quantization}} only for ideal, ballistic filaments), as well as from the effect of a small current flowing through the filament during the whole measurement time.}

    It was also found that resistive states corresponding to quantized conductance levels [an example is shown in Fig.~\ref{fig:samples}(c)] are less likely to remain stable as $G$ increases. In more detail, in the majority of experiments on the retention of the $1~G_0$ level, we observed well-stable states, which maintained the conductance for the whole recording time of 300~s. A similar situation was observed for the neighboring half-integer levels: $0.5~G_0$ and $1.5~G_0$. But the higher the $G$ value became, the more probable was a spontaneous change of the resistive state during the recording. In other words, an increase in $G$ corresponded to a decreased percentage of the states exhibiting quantized conductance values and stably maintaining them for 300~s. That is, the stability of levels corresponding to the $G = (n/2)\times G_0$ law decreases with $n$. {In the series of experiments that led to this conclusion}, for each of the conductance levels with $G/G_0 = 0.5, \; 1, \; 1.5, \; 2, \; 3, \text{ and } 5$, we measured 10~retention curves, taking into account each result---both positive (the states stored for at least 300~s within $\pm 0.2 G_0$) and negative (the states that changed during this time). All the parameters of the experiment remained unchanged throughout. The results obtained showed [see the inset in Fig.~\ref{fig:set-reset-quants}(c)] that, indeed, as $G$ increases, the proportion of the stable states corresponding to quantized conductance levels decreases, practically threefold (from roughly 60\% to 20\%).
    % {This observation prompted us to study in more detail the stability of resistive states from the statistical point of view, including an increase in the sampling, since only 10 measurements for each point were taken in these preliminary measurements}.

    Thus, we experimentally revealed that retention statistics could contain potentially interesting information on the properties of thin filaments with several conductance quanta. This inspired us to focus on a deeper experimental analysis of the retention statistics of the corresponding resistive states---{and, for this goal, to extend the preliminary experimental datasets of 10~measurements per $G$~point}---in search of nontrivial new signatures in the statistical distributions of stable states. In parallel, this triggered a preliminary theoretical analysis, which qualitatively suggested that one of such signatures could be an increased stability of thin filaments enabled by the electric current flowing through them. We will switch to the theory in Sec.~\ref{sec:Results_motivation}, ~\ref{sec:Results_simulation}, and now we report the methodology and results of our measurements of retention statistics in memristors.

    %\alert{To describe such a result, qualitative considerations were insufficient, and it was decided to try to describe the process by quantitative quantum-mechanical calculations. The ideas that were taken into account are outlined in the next section.}

    \subsection{Experimental stability of conductive filaments}
    \label{sec:Results_stability}

    As said, the retention properties of quantized conductance levels of the Cu/PPX/ITO memristors were measured in more detail. Now, for each of the six $G$ values equal to 0.5, 1, 1.5, 2, 3, and 5~$G_0$, the $G(t)$ dependence was recorded {with 20--30}~measurements for each $G$, for 300~s in steps of 1~s. As before, the initial state of the memristor was set using the ``write-verify'' algorithm~\cite{WriteVerify} with a precision of $0.1 G_0$, and a constant reading voltage $U_{\text{read}}$ was then applied to the memristor for the entire measurement time.

\begin{figure}[ht]
        \includegraphics[width=17cm]{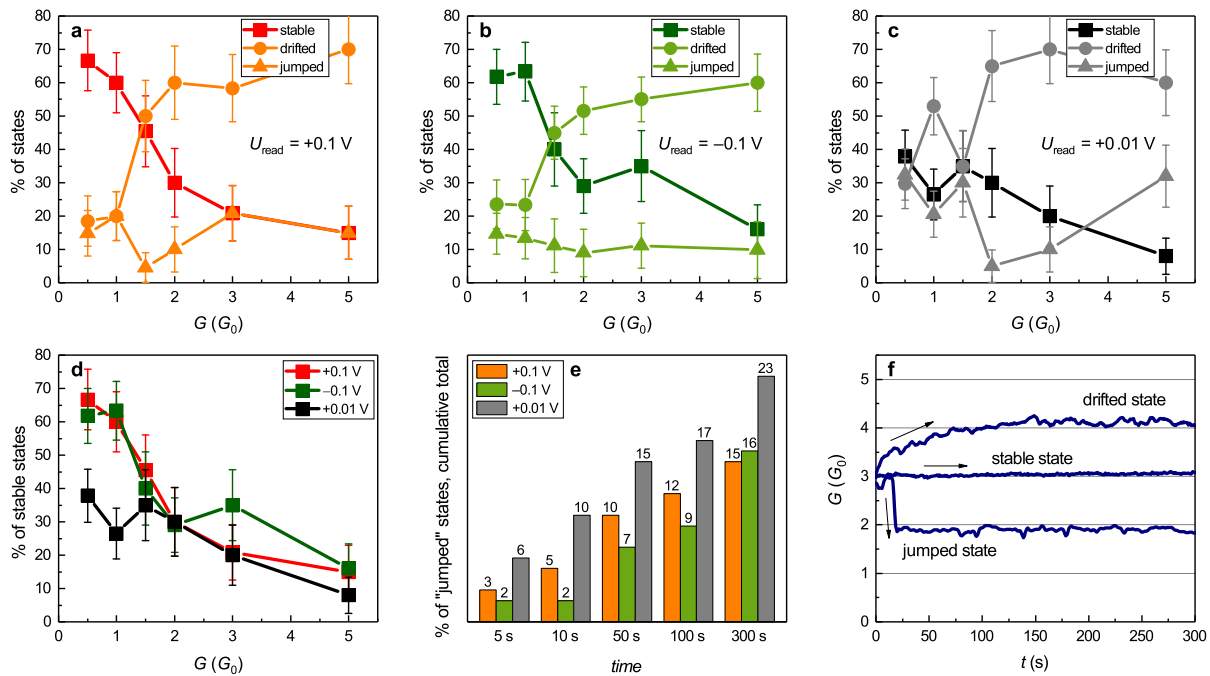}
        \caption{Stability of quantized conductance levels in Cu/PPX/ITO memristors. (a) Percentages of stable, drifted, and jumped states among the states at a preset conductance level $G$ for the reading voltage $U_{\text{read}} = +0.1~\text{V}$. (b) The same, but for $U_{\text{read}} = -0.1~\text{V}$. (c) The same, but for $U_{\text{read}} = +0.01~\text{V}$. (d) Percent of stable states versus $G$ for different reading voltages. (e) Cumulative {share} of jumped states over time for different reading voltages, {for all $G$~values taken together} (shown as bar heights and explicitly). (f) Examples of $G(t)$ dependences for stable, drifted, and jumped states.}
        \label{fig:stabil-exp}
    \end{figure}

    The obtained dependences were interpreted as follows. If the conductance of the memristor during the recording time changed within $\pm0.2~G_0$, such a state was considered stable. If the conductance changed more, the state was considered unstable. In the latter case, two variants were possible: smooth or abrupt changes of the conductance. On this basis, unstable states of the memristor were classified as either ``drifted'' (i.e., with a smooth change of $G$), or ``jumped'' (with a sharp change of $G$ by more than $0.5~G_0$). All the states, for which at least one conductivity jump was observed during the recording time, were considered as jumped. For example, Fig.~\ref{fig:stabil-exp}(f) shows $G(t)$ curves of the three types for $G = 3 G_0$: the stable state remained at the preset conductance level, while the unstable states changed (abruptly or smoothly) their conductance. Also, among the unstable states, we determined the percentages of those whose conductance decreased and of those whose conductance, on the contrary, increased. Thus, the tendency of the copper filament to rupture or, on the contrary, to thicken was determined, respectively.

    Initially, the $G(t)$ curves were recorded at the default reading voltage $U_{\text{read}} = +0.1~\text{V}$.  The measured dependences of the percentages of stable, drifted, and jumped states on $G$ are shown in Fig.~\ref{fig:stabil-exp}(a); {the error bars correspond to standard deviations of the multinomial distribution,  $\sigma = \sqrt{p(1-p)/N}$, where $p$ is the proportion of stable/drifted/jumped states obtained in a series of $N$ measurements.} One observes that for higher initial conductance levels, the proportion of stable states decreases {[which confirms the trend in the inset of \mbox{Fig.~\ref{fig:set-reset-quants}(c)} discussed in {Sec.~\ref{sec:Results_quantization}}]}. For instance, for $G = 0.5 G_0$ we observed {$(67 \pm 9)\%$} of stable states, but for $G = 5G_0$ there are only {$(15 \pm 8)\%$} of such states.
    %The number of both drifted and jumped states increases (from 3 to 5, i.e., by about 60\%, and from 1 to 3, i.e., by 200\%, respectively).
    The sign distribution of the conductance change among the unstable states turned out to be almost {symmetric (i.e., almost 50/50)}: the conductance increased for {$(53 \pm 6)\%$} of such states and decreased for the remaining {$(47 \pm 6)\%$ of unstable states. Finally,} among all the measurements {taken together}, a total of {$(42 \pm 4)\%$} of stable states were observed. We should note here that such a statistical study of stability of quantized conductance levels is, to the best of our knowledge, the first one existing in the literature. In all the available works presenting the corresponding retention curves, one does not mention, in which attempt these curves were recorded. In other words, it is {generally} assumed that the percentage of stable states is 100\%, which cannot be true even from a probabilistic point of view. In this regard, our study appears quite useful not only for testing theoretical models of RS and filament dynamics (including our theoretical argument presented in the below subsections), but also in the context of studying general properties of memristors exhibiting a conductance quantization effect.

    Next, to reveal the influence of the reading voltage on the stability of resistive states, we also carried out completely analogous series of measurements, but with different $U_{\text{read}}$ values. Fig.~\ref{fig:stabil-exp}(b) shows the dependences of the proportions of stable, drifted, and jumped $G(t)$ curves on the initial conductance $G$ for a negative reading voltage $U_{\text{read}} = -0.1~\text{V}$. Note that, in general, the dependences are quite similar to those shown in Fig.~\ref{fig:stabil-exp}(a), which indicates that there is little influence of the reading voltage sign on the stability of the retention curves. {In particular, the total percent of stable states was equal to $(43 \pm 4)\%$ at the negative reading voltage, i.e., remained practically unchanged compared with $(42 \pm 4)\%$ at $U_{\text{read}} = +0.1\text{ V}$ [see Fig.~\mbox{\ref{fig:stabil-exp}(a,b)}]. The complementary percent of unstable (drifted+jumped) states, of course, also remained virtually at the same level as before, but, notably, the decreased-conductance states now make up $(70 \pm 5)\%$ of all the unstable states, which is quite far from a 50/50 distribution, in contrast to the case of the positive reading voltage $U_{\text{read}} = +0.1\text{ V}$. Such an asymmetric behavior is expectable, indeed, since even at small negative currents, there is a tendency for reverse migration of copper ions towards the top electrode and, as a consequence, for the rupture of the conductive filaments.}

    We also carried out such measurements at a lower value of the reading voltage, namely, at $U_{\text{read}} = +0.01~\text{V}$: the results are shown in Fig.~\ref{fig:stabil-exp}(c). {One can observe that, in comparison with the two previous $U_{\text{read}}$ values, the percent of stable states is lower. The total share of stable states also decreased significantly to $(27 \pm 4)\%$.} This indicates that the current flowing through a thin filament can maintain its stability in some way, and, as its amplitude decreases, the stability of the filament also decreases. It is also interesting that at $U_{\text{read}} = +0.01~\text{V}$, the part of unstable states, which decreased their conductance, is {$(66 \pm 4)$\% (of the total number of unstable states), which is close to this figure} at $U_{\text{read}} = -0.1~\text{V}$.

    Observation of this unusual effect, certainly, calls for a theoretical explanation; a possible nature of the current-enhanced stability will be discussed in the next subsections. For an easier comparison, the dependences of the percentages of stable states on $G$ for different reading voltages are also shown in Fig.~\ref{fig:stabil-exp}(d). {In addition, a considerable statistical significance of the observed higher percentages of stable states for $|U_{\text{read}}| = 0.1~\text{V}$, as compared with those for $U_{\text{read}} = +0.01~\text{V}$, can also be justified by the $\chi^2$ analysis. Namely, for a comparison of our measurements at $U_{\text{read}} = +0.1~\text{V} \text{ and } +0.01~\text{V}$, the Yates-corrected $\chi^2$ test~{\cite{Yates}} gives the $p$-values below $0.05$ and below $0.02$ for $G = 0.5 G_0$ and $G = G_0$, respectively. In other words, the effect is justified with a $95\%$~probability for resistive states with the conductance of $0.5 G_0$ and with a probability over $98\%$ for those with $G = G_0$. For the measurements of both types of states taken together, the $\chi^2$ test gives $p \sim 0.006 = 0.6\%$. Analogously, comparing the other pair of reading voltages, namely, $U_{\text{read}} = -0.1~\text{V} \text{ and } +0.01~\text{V}$, we reveal the $p$-values below $0.08$ ($G = 0.5G_0$), below $0.01$ ($G = G_0$), and around $0.005$ ($G = 0.5 G_0$ and $G = G_0$ together), respectively. Roughly speaking, therefore, there is a $98.9\%$~probability that the current-enhanced stability effect takes place for the thinnest conductive filaments for $|U_{\text{read}}| = 0.1~\text{V}$ and both signs of this voltage. Below, we will discuss why such an effect may, indeed, show up especially for thin filaments with small conductance values.}

    {Apart from what was said above,} we believe that the jumped states determine the stability of the conductance level of the memristor better than the drifted states, because a jump change in conductance reflects a sharp and significant change in the geometry of the filament. Moreover, the timing of the jumps is just as important as their total number over the recording time. For example, it may turn out that most of the jumps occurred in the first 10 seconds of a $G(t)$ measurement, or, say, in the first 100 seconds. Obviously, the first case can be regarded as a much more unstable version of the filament. Fig.~\ref{fig:stabil-exp}(e) shows {the cumulative conductance jump statistics over different time windows at different reading voltages (with the measurements for all the $G$~values merged into a single dataset)}. It is clearly seen that at a small $U_{\text{read}} = +0.01~\text{V}$, the conductance jumps occur much earlier than at higher voltages. This observation also confirms the conjecture that the current flowing through the filament can stabilize it.

    In addition to the above results, we would like to note here the potential of statistical analysis of the stability of quantized conductance levels in memristors as a tool for studying the mechanism of RS in them. For example, though we previously established that RS in PPX-based memristors is driven by the ECM mechanism with a small number of conductive Cu filaments~\cite{ref16} (a conclusion which is quite sufficient for most applications), a study of retention statistics could be, in a sense, complementary to that, e.g., providing information on whether RS has to do with just a single conductive filament or, say, with two of them. A sketch on an argument revealing such information based on our retention statistics is presented in Appendix~\ref{app:probabilities} and can, in principle, be extended to strictly defined statistical criteria to be applied to the measured retention curves. Thus, in general, we can conclude that, despite the time and effort required for the measurements and data collection, statistical analysis of the retention properties is a potentially powerful tool for studying various aspects of RS in ECM/VCM memristors.

    \subsection{Current-enhanced filament stability: motivation of the theory}
    \label{sec:Results_motivation}

    In view of the {experimental signatures of the} enhanced stability of the lowest-conductance states in Cu/PPX/ITO memristors and a positive effect of the reading voltage on the filament stability, here we discuss a theoretical model of the effect of a quantum current through a filament on its shape and stability. We use a simple continuous model, in which the filament is described in terms of a spatial domain $\mathcal{D}$ with a certain profile, the conduction electrons being confined to it and exerting pressure on its walls (Fig.~\ref{fig:modelGeometry}). The corresponding second-quantized energy takes the form:
    \begin{equation}\label{H_model}
        \hat{H} = \frac{\hbar^2}{2m_\ast}\int_{\mathcal{D}} \bvec\nabla\hat\psi\hc \cdot \bvec\nabla\hat\psi \; \diff^3x
                    + \sigma \int_{\Sigma_{\text{lat}}} \diff{S}
                    +\lambda \left(\int_{\mathcal{D}} \diff^3x - V_0 \right),
                    \quad \left.\hat\psi\right\vert_{\Sigma_{\text{lat}}} = 0,
    \end{equation}
    where $m_\ast$ is the electron effective mass, $\hbar = h/2\pi$ is the reduced Planck's constant, $\hat\psi(\bvec{x})$
    is the electron field operator, and $\sigma$ is the surface energy per unit area of the filament lateral surface $\Sigma_{\text{lat}}$. The first and the second terms in $\hat{H}$ describe the free conduction electrons inside the filament and the surface tension, respectively. The third term fixes the total volume $V_0$ of the filament via a Lagrange multiplier $\lambda$; in what follows, we will also discuss the same model without the volume constraint, which formally corresponds to setting $\lambda = 0$.
    We choose a simple axisymmetric filament corresponding to $0 \le r \le R(z), \; 0 \le z \le L$ in cylindrical coordinates, with the asymptotic
    radii $R(0) = R(L) = R_0$, and an electron wave with zero angular momentum, i.e., independent of the azimuthal angle $\varphi$ (see Fig.~\ref{fig:modelGeometry}). For a fixed filament geometry, our model then determines a scattering problem on the wave function $\psi(r, z)$ of an electron with energy $E$ in the conducting channel $M = 1, 2, \ldots$:
    \begin{figure}[ht]
        \includegraphics[width=8.6cm]{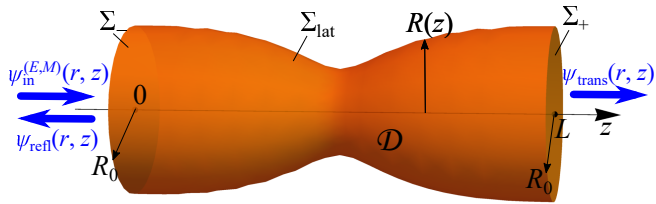}
        \caption{Geometry of the filament described by our theoretical model~\eqref{H_model}. Blue arrows represent the scattering boundary conditions~\eqref{boundaryConditions} featuring an incident wave $\psi^{(E,M)}_{\text{in}}$ of an electron with energy $E$ in the conducting channel $M$ and the reflected and transmitted waves $\psi_{\text{refl,trans}}(r,z)$.}
        \label{fig:modelGeometry}
    \end{figure}
    \begin{gather}\label{Schroedinger_eqn}
        -\frac{\hbar^2}{2m_\ast} \nabla^2 \psi(r,z) = E \psi(r,z), \qquad \psi(R(z), z) = 0,
        \\
        \label{boundaryConditions}
        \psi(r, z) = \sum\limits_{m = 1}^{\infty} \frac{J_0(\zeta_m r / R_0)}{R_0 \sqrt\pi |J_1(\zeta_m)|} \times
                    \begin{cases}
                        (\delta_{m,M} e^{\ii k_m z} + \mathfrak{r}_{m,M}(E) e^{-\ii k_m z}), & z\le 0, \\
                        \mathfrak{t}_{m,M}(E) e^{\ii k_m (z-L)}, & z \ge L,
                    \end{cases},
    \end{gather}
    where $J_\nu(x)$ are the Bessel functions of the first kind, $\zeta_m$ is the $m$th zero of $J_0$, and
    $k_m = \sqrt{2m_\ast E / \hbar^2 - \zeta_m^2 / R_0^2}$ is the channel wavenumber. After finding the amplitude transmission and reflection matrices $\mathfrak{t}_{m,M}(E), \mathfrak{r}_{m,M}(E)$, and thus the scattering wave function $\psi(r,z)$, one can find the ``quantum pressure'' exerted by a flowing current $I$ on the lateral surface of the filament (for details, see Appendix~\ref{app:Rmatrix}):
    \begin{gather}\label{quantumPressure}
        p(z) = \frac{e U}{2\pi} \sum_{M: \; k_M\in \Reals} \frac{1}{k_M} \left|\frac{\pd \psi(r,z | E, M)}{\pd n}\right|^2_{r = R(z)},
        \quad I = G_0 U \sum_{M: \; k_M\in \Reals} \mathcal{T}(M\to \text{all} | E),
        \\
        \mathcal{T}(M\to \text{all} | E) = \sum_{m: \; k_m \in \Reals} \frac{k_m}{k_M} |\mathfrak{t}_{m,M}(E)|^2,
    \end{gather}
    where the transmission probability $\mathcal{T}(M\to \text{all} | E)$ contains a sum over all open output channels. For simplicity, in this work we will use a monochromatic approximation with $E$ near the Fermi energy $E_{\text{F}}$; in general, one has to integrate over the energies and take a sum over all open channels (see Eq.~\eqref{p_integral}). The quantum pressure force is accompanied by the surface tension trying to minimize the area of the surface:
    \begin{equation}\label{surfaceTension}
        f(z) = -\frac{\sigma}{R(z)} \frac{1 + R'^2(z) - R(z) R''(z)}{(1+R'^2(z))^{3/2}}.
    \end{equation}
    In our calculations, we use the value of the surface energy $\sigma = 1.2~\text{N/m}$ for copper filaments from Ref.~\cite{WangIelmini_NatComm2019}
    (cf.~the value of about $1.7~\text{N/m}$ for the crystals~\cite{Copper_surfaceEnergy}). As for the Fermi energy, we take the bulk copper value $E_{\text{F}} = 7~\text{eV}$, along with the effective mass very close to the electron rest mass~\cite{Copper_EF_mast}. Of course, a continuous model~\eqref{H_model} can only be applicable semiquantitatively to thin filaments, but we believe it can still give us qualitatively correct information about the importance of the electron recoil (which we are calling the ``quantum pressure'' here) for the filament dynamics. An \textit{ab initio} study is far beyond the scope of the present work; instead, our simple model, containing a small number of parameters, is easier to analyze and readily gives estimations of the effects in question.

    Indeed, before resorting to numerical simulations of the quantum transport, it is instructive to estimate the forces driving the evolution of a filament with a single open channel ($n = 1$). In the case of a perfect cylindrical geometry with radius $R = \const$, the first channel opens for $R = R_{\min} = \zeta_1\sqrt{\hbar^2 / 2m_\ast E_{\text{F}}} \approx 1.8~\Angstrom$ (see Eq.~\eqref{boundaryConditions}).
    For such a channel supporting an ideal conductance of $G_0$, the surface tension on its walls is $f = -\sigma / R \sim -0.04~\text{eV}/\Angstrom^3$.
    On the other hand, at the voltage $U = 1~\text{V}$, such a channel supports a current $I = G_0 U = 7.7\times 10^{-5}~\text{A}$ corresponding to $\dot{N}_e \sim 0.5\times 10^{15}$ electrons per second passing through;  these electrons also exert a pressure $p = 0.06~\text{eV}/\Angstrom^3$, as given by Eq.~\eqref{quantumPressure} with $k_M = \sqrt{2 m_\ast (E + |e|U) / \hbar^2 - \zeta_1^2 / R^2}$. Though this outward pressure even turns out to dominate the surface tension forces for $R = R_{\min}$, a channel with $R = 1.5 R_{\min} \approx 2.7~\Angstrom$ has a surface tension $f \sim -0.03~\text{eV}/\Angstrom^3$ stabilizing the effect of a much weaker quantum pressure $p \sim 0.005~\text{eV}/\Angstrom^3$. Moreover, for $R = R_{\min}$ but a lower voltage, the surface tension can clearly counteract the quantum pressure. In any case, it turns out that the two effects may be competing in the typical range of parameters pertaining to memristor operation.

    The above picture may drastically change for a channel having a ``defect'', e.g., in the form of a constriction (see Figs.~\ref{fig:modelGeometry}, \ref{fig:simulation_filamentShapes}), which also prevents the electrons from a fully ballistic transport with $\mathcal{T} = 1$. Such a constriction results in a nontrivial reflected wave interfering with the incident one (see Eq.~\eqref{boundaryConditions}) and perturbing the pressure~\eqref{quantumPressure}. The interference pattern is expected to wash out at distances $\gtrsim 1/k_{\text{M}} \sim 1~\Angstrom$ from the constriction, thus, the total force on the filament due to electron recoil $F_z =  \frac{1 - \mathcal{T}}{\mathcal{T}} \times 2 \hbar k_M \dot{N}_e$, roughly speaking, acts on a surface area $\Delta{S} \sim 2\pi R / k_M$ around the constriction. The resulting pressure, which can be estimated as $\Delta{p} \sim F_z / \Delta{S}$, equals $0.05\frac{1 - \mathcal{T}}{\mathcal{T}}~\text{eV}/\Angstrom^3$ for $R = 1.5 R_{\min}$ and could even dominate the surface tension contribution $f$ in the narrowest parts of wider channels. Based on these rough estimations, we thus expect that (i) the shape of thin conducting filaments should be sensitive to the recoil of conduction electrons, (ii) the recoil, or quantum pressure forces should be localized near the narrowest segments of the filament and strongly depend on the transmission probability $\mathcal{T}$. Namely, for ballistic filaments with $\mathcal{T} = 1$, the total recoil force $F_z$ vanishes, which suggests its possible role in the stability of channels with quantized conductance values. Let us now check these expectations using numerical solutions of the scattering problem~\eqref{Schroedinger_eqn}.

    \subsection{Current-enhanced filament stability: numerical simulations}
    \label{sec:Results_simulation}

\begin{figure}[ht]
        \includegraphics[width=14cm]{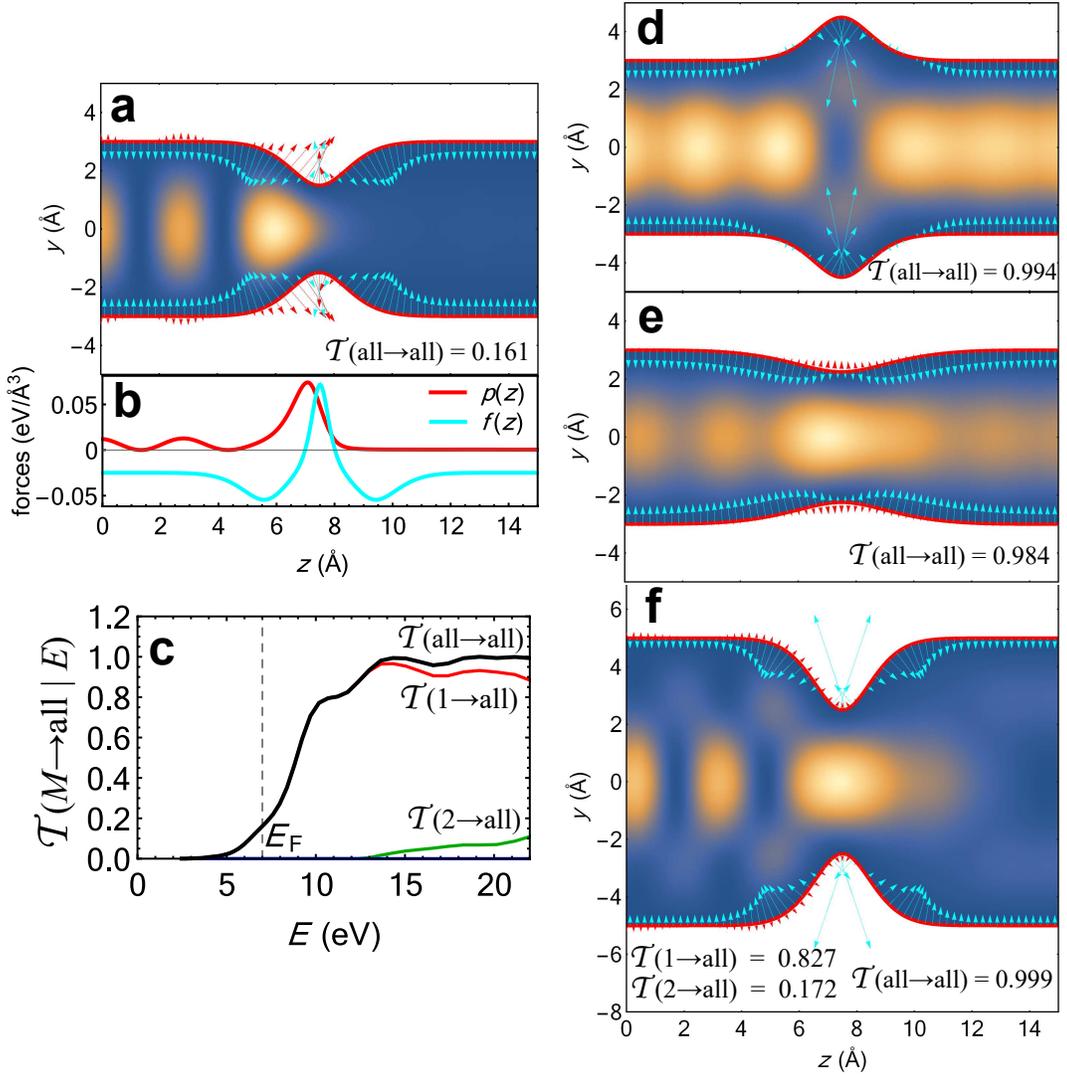}
        \caption{Quantum transport in filaments of different profiles and the forces acting on their lateral surfaces. Panels~(a), (d), (e), (f) show the probability density distribution $|\psi(r,z)|^2$ in a cross section of the filament, the resulting quantum pressure field $p(z)$ (red arrows), and the surface tension field $f(z)$ (cyan arrows). In these panels, the units of the force fields are the same, the voltage $U = 1~\text{V}$, and the wave functions are calculated at the Fermi surface $E = E_{\text{F}} = 7~\text{eV}$. Panel~(b) depicts the fields $p(z)$ and $f(z)$ as functions of $z$. Panel~(c) demonstrates the dependence of the transmission probabilities $\mathcal{T}(M\to\text{all} | E)$ on the charge carrier energy for filament~(a).
        }
        \label{fig:simulation_filamentShapes}
    \end{figure}

    Since a direct solution of Eq.~\eqref{Schroedinger_eqn} as an initial-value problem along $z$ (the so-called matching method) is known to be poorly-conditioned~\cite{Qtransport_conditioning1, Baye_Rmatrix_RPP2010, Qtransport_conditioning2}, we used the R-matrix method instead~\cite{Baye_Rmatrix_RPP2010, Wulf2020} (for details, see Appendix~\ref{app:Rmatrix}). Let us start from an analysis of the quantum pressure and surface tension distributions on surfaces of constrictions with different shapes and radii (Fig.~\ref{fig:simulation_filamentShapes}). The filament cross-sections in this figure demonstrate the density $|\psi(r,z)|^2$ of the electron wave at the Fermi energy, along with the quantum pressure field $p(z)$ and the surface tension $f(z)$; the current necessary for the evaluation of $p(z)$ was calculated as $I = \mathcal{T} G_0 U$, with the applied voltage $U = 1~\text{V}$ of the order of the RS voltage for the studied memristors [see Fig.~\ref{fig:samples}(b)]. Fig.~\ref{fig:simulation_filamentShapes}(a,b) clearly reveals that near a tight constriction, reflection occurs, resulting in an interference of the incident and the reflected waves and in a peak of the quantum pressure that competes with the surface tension force. Interestingly, the surface tension depends on the local mean curvature of the filament surface (see Eq.~\eqref{surfaceTension}), whereas the quantum pressure is nonlocal and, due to diffraction, is virtually insensitive to the defects much smaller than the inverse channel wavenumber $1/k_M$. The quantum pressure within our model always acts outwards (for both current directions!) and thus produces a stabilizing effect on the filament, which could, in principle, be torn apart by the surface tension forces. Note that we have intentionally chosen the filament geometry in Fig.~\ref{fig:simulation_filamentShapes}(a) to hit a slope of the transmission probability [see Fig.~\ref{fig:simulation_filamentShapes}(c)] with $\mathcal{T} \approx 0.16$, in order to analyze what could happen to a thin filament with an open, but non-ballistic conductive channel.

    Interestingly, the opposite, convex filament case depicted in Fig.~\ref{fig:simulation_filamentShapes}(d), in which the transport is virtually ballistic, reveals a different picture: the small ``bump'' on the filament profile turns out to be in the geometrical shadow zone, the electrons not getting there and not exerting any pressure on its walls. The surface tension, however, is still there and tends to pull the bump back towards the axis of the filament.  Moreover, the ballistic value of $\mathcal{T} \approx 0.994$ is likely to be directly related to the fact that the bump lies in the shadow zone for the electrons. Further, Fig.~\ref{fig:simulation_filamentShapes}(e) demonstrates that the quantum pressure force is also able to protect smoother constrictions in the places where the surface tension could otherwise dominate, leading to further deformation of the filament. Finally, Fig.~\ref{fig:simulation_filamentShapes}(f) shows that the effect should be suppressed for thicker filaments. In all the cases, one observes that the effect of the recoil of charge carriers is favorable for maintaining ballistic transport in thin filaments and for deforming the profiles of the non-ballistic ones. That is, the tentative outcomes of our semiqualitative analysis in Sec.~\ref{sec:Results_motivation} seem to be correct.

    Let us now discuss the effects the quantum pressure may have on the evolution of the filament profile. Leaving a full dynamical theory beyond the scope of the present study, we simply optimize the filament profile $R(z)$ using gradient descent iterations under the action of the pressure and surface tension forces. As mentioned above, we consider two cases, with and without the volume constraint $\int_0^L \pi R^2(z) \diff{z} = V_0$, and in the first case, the iterations are performed along the constraint surface:
    \begin{gather}\label{gradDescent}
        R(z) \to R(z) - \alpha g(z)R(z), \\
        g(z) = \begin{cases}
                    -f(z) - p(z) + \frac{\pi}{V_0} \int_0^L (f(z') + p(z')) R(z')\diff{z'}, & V = V_0 = \const, \\
                    -f(z) - p(z), & \text{no volume constraint},
                \end{cases}
    \end{gather}
    where $\alpha > 0$ is the learning rate.

    \begin{figure}[ht]
        \includegraphics[width=14cm]{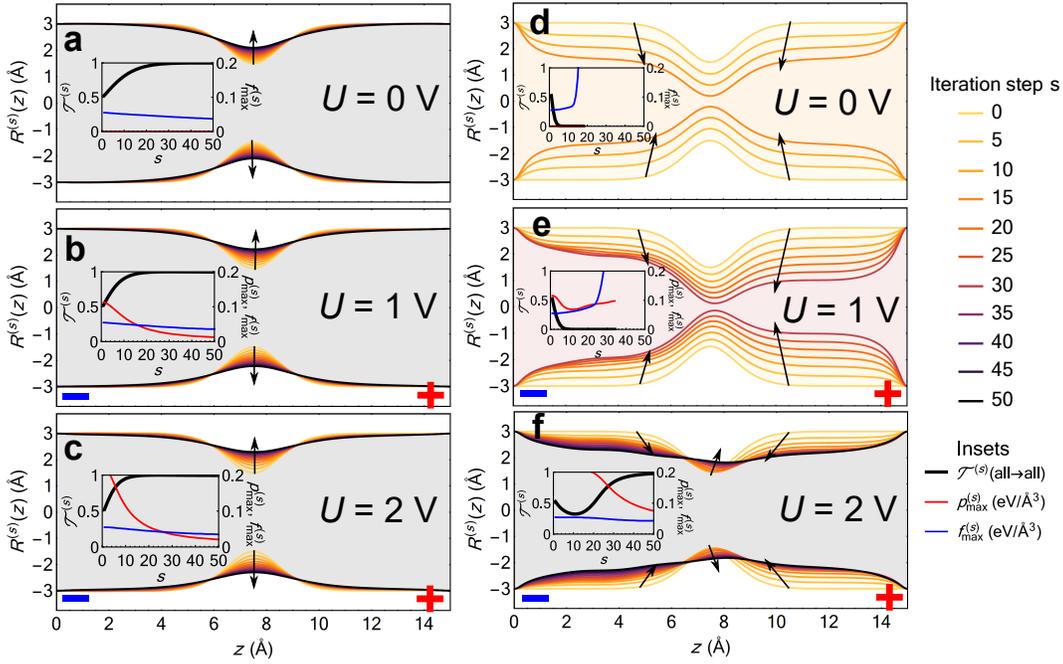}
        \caption{Filament profile iterations at different voltages: (a), (b), (c) with fixed total volume of the filament and (d), (e), (f) without the volume constraint. The learning rate $\alpha$ [see Eq.~\eqref{gradDescent}] is the same in all the six panels and is kept constant during the iterations. The insets show the changes of the transmission probability $\mathcal{T}(E)$, the maximum absolute values of the quantum pressure and the surface tension, $p_{\text{max}}, f_{\text{max}}$, in $\text{eV}/\Angstrom^3$, as the channel deforms. The direction of deformation is also shown with the arrows. For a better comparison of the effects, we took the same energy $E = E_{\text{F}} + 2~\text{eV} = 9~\text{eV}$ in all the panels.
        }
        \label{fig:simualtion_filamentEvolution}
    \end{figure}

    The results of our simulations are presented in Fig.~\ref{fig:simualtion_filamentEvolution} for different voltages $U = 0, \; 1, \text{ and } 2~\text{V}$; the learning rate $\alpha$ in Eq.~\eqref{gradDescent} is the same in all the panels and is fixed during the iterations, to provide qualitative visual information on the filament deformation rates. The evolution of the filament with a fixed volume [Fig.~\ref{fig:simualtion_filamentEvolution}(a,b,c)] may appear not very interesting at first glance: the gradient descent remains driven mainly by the surface tension most of the time, and the visual effect of the quantum pressure on the evolution of the filament profile seems limited to a modest acceleration of its ``smoothening''. However, a closer look at the insets in Fig.~\ref{fig:simualtion_filamentEvolution}(b,c) reveals that the quantum pressure does quite dominate the surface tension during the first iterations, when the transmission probability is not yet close to unity; as we saw in Fig.~\ref{fig:simulation_filamentShapes}, it pushes the constriction outwards, trying to make the filament smoother (together with a yet weaker surface tension force). After the conditions for ballistic transport are approximately achieved ($\mathcal{T} \sim 1$), the quantum pressure drops down, and the geometry of the filament continues to smoothen because of the surface tension. However, during these, later iterations, it is only the filament profile that is evolving, while the transmission probability remains constant, very close to unity. Thus, in terms of conductance quantization, the effect of the quantum pressure is much more favorable for its emergence than it seems to be; in fact, the evolutions of the transmission probability $\mathcal{T}^{(s)}$ [see insets in Fig.~\ref{fig:simualtion_filamentEvolution}(a,b,c)] evidence that the pressure-induced boost towards the ballistic regime is about $100\%$ at the voltages $U=1\text{--}2\text{ V}$. Interestingly, the quantum pressure is also ``economic'' in that it persists no more once the ballistic transport is achieved.

    If we lift the volume constraint, the picture changes drastically, see Fig.~\ref{fig:simualtion_filamentEvolution}(d,e,f). Indeed, in this case, at zero voltage, the evolution of the filament profile is nothing but a minimization of its surface area, which leads to a catenoid; such minimal surfaces, however, exist only for $L/2R_0 < 0.66$~\cite{Catenoid}. In our case, $L \gg 2R_0$, and the filament is rapidly torn apart by the surface tension, that is, the transmission probability evolves towards zero [Fig.~\ref{fig:simualtion_filamentEvolution}(d)]. With the applied voltage, the destructive evolution of the filament becomes more complicated, and one observes a spectacular change of the behavior between $U = 1~\text{V and } 2~\text{V}$: in the first case, the filament is still unstable, while for the higher voltage, the quantum pressure takes a lead, bringing the filament profile closer to that of a cylinder and the transmission to a ballistic one with $\mathcal{T} \approx 1$.

    The filament profile optimizations in Fig.~\ref{fig:simualtion_filamentEvolution} indicate that at realistic RS voltages, the quantum pressure force can, indeed, have an important role in the formation and evolution of thin filaments with quantized conductance. On the other hand, the enhanced stability effects observed in our experiments deal with voltages $U\sim 0.1~\text{V}$ one order of magnitude lower, {so that, at first glance, it may seem that the quantum pressure can hardly be relevant for these effects. However, a closer look reveals that the effects of the quantum pressure should be quite important even at the lower voltages. Indeed, e.g., {Figs.~\ref{fig:simulation_filamentShapes}(b),~\ref{fig:simualtion_filamentEvolution}(b,e)} indicate that the quantum pressure at $U = 1\text{ eV}$ is around $\text{25--50} \text{ meV}/\Angstrom^3$, and, since it is proportional to the voltage (see Eq.~{\eqref{quantumPressure}}), one estimates it at the level of $\text{2.5--5} \text{ meV}/\Angstrom^3$ at $U = 0.1\text{ eV}$, which cannot overtake the surface tension forces. However, Fig.~{\ref{fig:simualtion_filamentEvolution}} demonstrates a \emph{``deterministic''}, fast deformation of the filament, resembling a forced resistive switching event, while a much slower, \emph{stochastic} deformation, whose straightforward simulation is hardly feasible within the scope of the present work, should be governed by energy barriers and thermally activated motion of individual copper atoms/ions~{\cite{WangIelmini_NatComm2019}}. Now, an additional quantum pressure of $\Delta{p} = \text{2.5--5} \text{ meV}/\Angstrom^3$ implies an additional energy density, which, for fcc copper with the atom number density $n_{\text{Cu}} \approx 9\times 10^{22}\text{ cm}^{-3}$, corresponds to an extra energy $\Delta{E} \sim \Delta{p} / n_{\text{Cu}} \sim \text{25--50}\text{ meV} \sim (\text{1--2})~kT$ per one copper atom, or roughly $\Delta{E}_{\text{dof}}(0.1~\text{V}) = 0.5~kT$ per degree of freedom ($k$ is the Boltzmann constant, $T = 300\text{ K}$). For such a modification of the activation energies preventing the filament atoms from migration, one may expect that the probabilities of filament conductance jump/drift events can change by as much as $1 - e^{ -( \Delta{E}_{\text{dof}}(0.1\text{ V}) - \Delta{E}_{\text{dof}}(0.01\text{ V}) ) / kT } \sim 35\%$ between the reading voltages of $0.1~\text{V}$ and $0.01~\text{V}$, which is qualitatively consistent with our experimental observations for $G/G_0 = 0.5 \text{ and } 1$. Moreover, the magnitude of the quantum pressure effect at $U = 0.01~\text{V}$ is expected to be considerably smaller, $1 - e^{-\Delta{E}_{\text{dof}}(0.01\text{ V}) / kT} < 5\%$, which is below the statistical errors of our experiments.}

    {Therefore, without providing an atomistic simulation of the filament evolution within the present study}, we still expect the quantum pressure to be able to contribute to stability of thin filaments at the level of tens of percent. There are also other factors that appear to be important for the evolution of atomic-scale filaments, e.g., the surface energy can, in principle, be way smaller than its macroscopic values; {also, at higher voltages $U \gtrsim 1\text{ V}$, charge redistribution between copper atoms and ions should be accounted for, coupling the atomistic dynamics with electrochemical kinetics.} The effect of the nontrivial distribution of the electrostatic potential was not included in our simple model either: in fact, the corresponding forces acting on the copper ions in a strong electric field may compete with the recoil effects discussed by us here. These effects are clearly to be included into the simulations in further, generalized versions of the model.  Thus, strictly speaking, the recoil nature of the observed current-enhanced stability effect does not completely follow from the present model, requiring a multiscale atomistic simulation, such as the one in Ref.~\cite{WangIelmini_NatComm2019}, to decide; however, our experimental observations have revealed certain unusual features of the effect (such as its presence for both signs of the reading voltage, see Sec.~\ref{sec:Results_stability}), which are also predicted by our theoretical model. At the same time, this model provides a strong, semiquantitative indication that the recoil effects may be important for RS in the thin-filament regime, and should be taken into account in more sophisticated microscopic models of memristive devices.

    \section{Conclusions}
    \label{sec:Conclusion}
     We have thus studied the effects of retention statistics in Cu/PPX/ITO memristors and suggested a theoretical framework for analyzing the evolution of thin filaments. On both sides of the research, we have observed a somewhat counterintuitive effect: a raised reading voltage stabilizes the resistive state. Indeed, at first glance, both due to Joule heating effects and to copper ion migration, the applied voltage is likely to destroy the filaments or make them grow. In contrast, we have observed that even a negative reading voltage $U_{\text{read}} = -0.1~\text{V}$, which is meant to pull the memristor towards its off-state, helps it to maintain its on-state with a quantized conductance. {Moreover, for both voltage signs, $U_{\text{read}} = \pm0.1~\text{V}$, the effect is especially pronounced for the lowest-conductance states of the memristor; our measurements justify its existence for these states with a 99\% significance.} In fact, it was an idea coming from an order-of-magnitude estimation of the recoil effects (see Sec.~\ref{sec:Results_motivation}) that served as a motivation for the experimental analysis of retention statistics at different voltages, including the negative ones. Believing that this could not be just a coincidence, using both the theory and the experiment, we have further found a strong indication of the importance of the effects studied in the process of RS. In particular, we have demonstrated that at voltages typical for RS, the geometries of thin filaments become driven by the quantum pressure forces, due to the recoil of flowing charge carriers, and these forces, together with the surface tension, tend to set the transmission probability either to $\mathcal{T} = 0$, or to the ballistic one, $\mathcal{T} = 1$.
     %[see Fig.~\ref{fig:simualtion_filamentEvolution}(d,e,f)].
     It seems interesting then to study retention statistics under different voltages in other types of memristors, possibly matching the existence of the current-enhanced stability effect with the clearly observed quantized conductance levels in these devices. Another direction that has to be further followed is the simulation of electron recoil within an atomistic, possibly, multiscale \textit{ab~initio}--molecular dynamics framework. This clearly is a more challenging study, for which, however, the present work can feature as an inspiring proof of principle. On the experimental side, such research can be fuelled by our new technique of analyzing the retention statistics in different conditions of the memristor operation. Working in a ``non-invasive'', non-disturbing way considerably below the RS voltage, this technique is able to extract important data about the formation, evolution, and decay of conductive filaments, and, consequently, about the RS effect. We hope that this technique and the new possible mechanism of filament stability we have suggested in this work will pave the way toward a better understanding of RS and retention of resistive states in memristors and will eventually open up new opportunities of their rational design.

    \section{Samples and methods}
    \label{sec:Methods}

    PPX-based memristors were fabricated using the following method. {The commercially purchased $20\times20~\text{mm}^2$ ITO-coated glass (the bottom electrode, with the ITO thickness around $50~\text{nm}$) was used as a substrate, and the parylene-N layers ($\sim 100~\text{nm}$ in thickness) were deposited on it by the gas-phase surface polymerization method using an SCS Labcoater PDS~2010 vacuum deposition system. PDS~2010 transforms a parylene dimer (2,2-para-cyclophane or its derivatives) into a gaseous monomer; the material polymerizes on the substrate by deposition at room temperature. The method provides a uniform polymer application on all sides of the substrate, resulting in a truly conformal coating. To improve adhesion and ensure a uniform PPX coating, the substrates were thoroughly cleaned of oils and solvents beforehand.} After that, the top $0.2 \times 0.5~\text{mm}^2$ Cu~electrodes with a thickness of $\sim 500~\text{nm}$ were
    deposited by magnetron sputtering through a shadow mask using a Torr MagSput~4G2-RF/DC deposition system. As a result, we obtained an array
    of single Cu/PPX/ITO memristors. Structural investigations were carried out with the transmission electron microscope Titan 80-300 (FEI, USA). The cross-sectional preparation of the memristive samples was done by the focused ion beam method on Helios 600i. Memristive characteristics of the devices were measured using a Cascade Microtech~PM5 analytical
    probe station; the voltage pulses were supplied by a Keithley~2636B SourceMeter, programmed in LabView. All experiments were
    performed at room temperature.

    \section*{Acknowledgments}
     The authors are grateful to Dr.~A.~E.~Kazantsev (The University of Manchester), Dr.~A.~V.~Emelyanov, and Dr.~V.~A.~Demin (NRC ``Kurchatov institute'') for fruitful discussions. The authors are also grateful to A.~A.~Nesmelov and K.~Yu.~Chernoglazov (NRC ``Kurchatov institute'') for the assistance in samples preparation and to E.~V.~Kukueva (NRC ``Kurchatov institute'') for the electron microscopy. The work has been partially supported by
    the Russian Foundation for Basic Research (project No.~20--07--00696) in the part concerning quantized conductance stability measurements and the theoretical model, and by the Russian Science Foundation (project No.~20-79-10185) in the part concerning general characterization of the samples (I-V curves, retention, microscopy). All the measurements have been carried out with the equipment of the Resource Centers of National Research Center ``Kurchatov Institute''.

    \vspace{1em}

    The contributions of the authors are as follows: O.~K. developed the theoretical model and performed the quantum transport simulations; A.~M. and B.~S. conducted the measurements; A.~M. processed the experimental data; A.~M., O.~K., and V.~R. conceptualized the study; A.~M. directed its overall design; O.~K., A.~M., and V.~R. prepared the manuscript.

    \appendix

    \section{Retention statistics and single versus multiple conductive filaments}
    \label{app:probabilities}

    Here we give an example of how a statistical analysis of conductance jumps (see Sec.~\ref{sec:Results_stability}) can provide information on the nature of RS, namely, favor a single-filament RS mechanism versus several thin filaments, in the quantized-conductance regime. Leaving an elaborate analysis, including formulation of the statistical criteria, for a future study, in this work we limit ourselves to a simple argument distinguishing between the two hypotheses. Let us start by denoting the conductance lifetime of a single filament with $G = n G_0$ (i.e., the average storage time of its conductance within $\pm 0.2 G_0$) as $\tau_n$. Then, for a \emph{single-filament} RS mechanism, we expect an exponential decay of the percent of stable states with conductance $G = n G_0$: $w_{\text{stable}}(t | n G_0) \sim e^{-t / \tau_n}$. If, on the contrary, a resistive state $G = n G_0$ of the memristor is represented by, say, \emph{two filaments} with conductances $n_1 G_0$ and $n_2 G_0$, \; $n_1 + n_2 = n$, \; the stability of this state against RS is expected to obey the law $w_{\text{stable}}(t | n_1 G_0 + n_2 G_0) \sim e^{-t /\tau_{n_1}} \times e^{-t / \tau_{n_2}}$: for a stable state, neither of the two filaments should undergo an instability. Roughly substituting $n_1 = n_2 = n/2$, we observe that $w_{\text{stable}}(t | nG_0) \sim e^{-2t / \tau_{n/2}} \sim w_{\text{stable}}^2(t | nG_0/2)$, i.e., should fall off exponentially with the conductance. Such a fast decay law is very unlikely in the case of a single filament, and, in fact, Fig.~\ref{fig:stabil-exp}(d) does not exhibit it. Similarly, one can analyze the features of the number of jumped states in Fig.~\ref{fig:stabil-exp}(a,b,c): now, a multifilament character of RS favors an exponential dependence of $1 - w_{\text{jumped}}(t | nG_0)$ on $n$, {which is not what is actually observed in {Fig.~\ref{fig:stabil-exp}}}. That is, we can conclude that we were most likely observing a single filament in our devices. We thus believe that the above statistical technique is a potentially powerful tool for studying the RS mechanisms, especially interesting for memristors with quantized conductance levels and probably not restricted solely to Cu/PPX/ITO devices, but applicable to other types of memristors with a filamentary switching mechanism.

    \section{Quantum transport through a filament in the R-matrix formalism}
    \label{app:Rmatrix}

    The R-matrix method we used for the determination of the electron scattering wave function $\psi(r,z)$ in a filament with a profile $R(z)$ (see Eqs.~\eqref{Schroedinger_eqn}, \eqref{boundaryConditions}) is quite customary in quantum transport simulations~\cite{WignerEisenbud, Wulf2020}. This method makes use of a linear integral equation relating the value of the wave function in the scattering domain $\mathcal{D}$ (i.e., for $z\in (0,\; L)$, see Fig.~\ref{fig:modelGeometry}) to its normal gradients at the `leads' $\Sigma_\pm$ (i.e., at $z = 0,\; L$):
    \begin{equation}\label{Rmatrix_integralEquation}
        \psi(r,z) = \int_{0}^{R_0}\left[ \mathcal{R}(r,z; r',L; E) \pd_z\psi(r',L) - \mathcal{R}(r,z; r',0; E) \pd_z\psi(r',0) \right] \times 2\pi r'\diff{r'},
    \end{equation}
    {where $\psi$ is assumed to obey the Schroedinger equation in $\mathcal{D}$ and the Dirichlet boundary conditions on the lateral surface $\Sigma_{\text{lat}}$ (see Eq.~{\eqref{Schroedinger_eqn}}). The R-matrix is conventionally expanded in a series over the so-called Wigner--Eisenbud functions $\chi_\nu(r,z)$, an
    orthonormal set of eigenfunctions of $-\hbar^2\nabla^2 / 2 m_\ast$ obeying the Neumann boundary conditions at the leads $\Sigma_\pm$ and the Dirichlet ones on $\Sigma_{\text{lat}}$, with the eigenvalues $E_\nu$~{\cite{WignerEisenbud}}. It is straightforward to show, indeed, that with these combined boundary conditions, $-\nabla^2$ is a Hermitian operator allowing for a complete set of eigenfunctions in $\mathcal{D}$. Now, the expansion coefficients of $\psi$ over this set can be revealed from the following expression {\cite{Wulf2020}}:}
    \begin{eqnarray}
        \frac{2 m_\ast(E_\nu - E)}{\hbar^2} \int_{\mathcal{D}} \chi_\nu\cc \psi \;\diff^3x
        &=& \int_{\mathcal{D}} (\chi_\nu\cc \nabla^2 \psi - \nabla^2\chi_\nu\cc \cdot \psi) \;\diff^3x \nonumber \\
        &=& \oint_{\pd\mathcal{D}} \left(\chi_\nu\cc \frac{\pd\psi}{\pd n} - \frac{\pd\chi_\nu\cc}{\pd n} \psi \right) \;\diff{S} \nonumber \\
        &=& \int_{\Sigma_+ \cup \Sigma_-} \chi_\nu\cc \frac{\pd\psi}{\pd{n}} \;\diff{S},
    \end{eqnarray}
    {where the surface integral has reduced to the two bases $\Sigma_\pm$ because of the boundary conditions on $\chi_\nu$ and $\psi$. The above relation between the solution $\psi$ and its normal derivatives $\pd\psi/\pd{n}$ on $\Sigma_\pm$ agrees with Eq.~{\eqref{Rmatrix_integralEquation}} and finally yields a series expansion for the R-matrix itself:}
    \begin{equation}
        \mathcal{R}(r,z; r',z'; E) = \frac{\hbar^2}{2 m_\ast} \sum_{\nu} \frac{\chi_\nu(r,z) \chi_\nu\cc(r',z')}{E_\nu - E}.
    \end{equation}
    Next, taking Eq.~\eqref{Rmatrix_integralEquation} for $z = 0, L$ and substituting the boundary conditions~\eqref{boundaryConditions} into it, one arrives at two matrix equations on the complex $\mathfrak{t}_{m,M}(E)$ and $\mathfrak{r}_{m,M}(E)$ coefficients,
    \begin{gather}
        (\idMatrix - \mathds{R}(E)\cdot\mathrm{i}\mathds{k})\twoCol{\mathfrak{r}}{\mathfrak{t}} =
        -(\idMatrix + \mathds{R}(E)\cdot\mathrm{i}\mathds{k})\twoCol{\idMatrix}{\mathbb{0}}, \qquad \mathds{k} \equiv
        \diag(k_1, k_2, \ldots),
        \\
        \twoMatrix{\mathcal{R}(r,0;r',0;E)}{\mathcal{R}(r,0;r',L;E)}{\mathcal{R}(r,L;r',0;E)}{\mathcal{R}(r,L;r',L;E)} \equiv
        \frac{1}{\pi R_0^2} \sum_{m,m'=1}^\infty \mathds{R}_{m,m'}(E) \frac{J_0(\zeta_m r / R_0)}{|J_1(\zeta_m)|}
        \frac{J_0(\zeta_{m'} r' / R_0)}{|J_1(\zeta_{m'})|}.
    \end{gather}
    After fixing the total number of channels to $m_{\text{max}} = 10$ and solving these equations, we revealed the wave function in the whole scattering domain again by virtue of Eq.~\eqref{Rmatrix_integralEquation}.

    The expression~\eqref{quantumPressure} for the pressure can be found directly from the action corresponding to the electronic part of the energy~\eqref{H_model}:
    \begin{equation}
        S_{\psi} = \int
                    \left\{ \frac{\ii\hbar}{2}(\psi\hc \dot\psi - \dot{\psi}\hc \psi)
                           -\frac{\hbar^2}{2m_\ast} \bvec\nabla\psi\hc \cdot \bvec\nabla\psi
                    \right\}
                    \diff^3x \diff{t}.
    \end{equation}
    Namely, the normal-normal component of the stress tensor, related to the pressure, reads
    \begin{equation}\label{T_nn}
        T_{nn} = \frac{\hbar^2}{2m_\ast} \left| \frac{\pd\psi}{\pd n} \right|^2,
    \end{equation}
    when evaluated on the lateral surface $\Sigma_{\text{lat}}$ with $\psi = 0$. To obtain the pressure, it remains to fix the normalization of the wave function so that $|\psi|^2$ corresponds to the electron number density (note that Eq.~\eqref{boundaryConditions} uses another normalization convention, which is common in scattering problems). Indeed, a current $I$ through the $M$th channel of the filament corresponds to $I / |e|\mathcal{T}$ incident electrons per second, however, Eq.~\eqref{boundaryConditions} gives the incident particle number current equal to the channel velocity $\hbar k_M / m_\ast$. Inserting the missing coefficient
    $I m_\ast / |e| \hbar k_M \mathcal{T}$ in front of the stress tensor~\eqref{T_nn} and noting that $I = G U = \mathcal{T} G_0 U$, one yields Eq.~\eqref{quantumPressure} for the pressure. In fact, one can also derive a general expression for the quantum pressure within the Landauer--B\"uttiker formalism \cite{ref14, Buttiker}, assuming two reservoirs with chemical potentials $\mu_{\text{L}} = \mu_{\text{R}} + eU$ and $\mu_{\text{R}} < \mu_{\text{L}}$ at the left and the right leads, respectively; we only quote the result here:
    \begin{equation}\label{p_integral}
        p(z) = \int_{\mu_{\text{R}}}^{\mu_{\text{L}}} \diff{E} \sum_{M: \; k_{M}(E) \in \Reals} \frac{1}{2\pi k_M(E)}
                  \left| \frac{\pd \psi(r,z | E, M)}{\pd n}\right|^2_{r = R(z)}.
    \end{equation}
    This result, in particular, takes the form of Eq.~\eqref{quantumPressure} for $eU \to 0$.

\end{document}